\documentclass[multphys,vecphys]{svmult}

\usepackage{makeidx}         
\usepackage{graphicx}        
\usepackage{multicol}        

\makeindex             

\begin{document}

\title*{Young Kinematically Decoupled Components in Early-Type Galaxies}
\author{Richard M. McDermid\inst{1}, Eric Emsellem$^{2}$, Kristen L.\ Shapiro$^{3}$, R. Bacon$^{2}$, M. Bureau$^{4}$, Michele Cappellari$^{1}$, Roger L.\ Davies$^{4}$, P.~T.\ de Zeeuw$^{1}$,\\ Jes{\'u}s Falc{\'o}n-Barroso$^{1}$, Davor Krajnovi{\'c}$^{4}$, Harald Kuntschner$^{5}$,\\ Reynier F.\ Peletier$^{6}$, \and Marc Sarzi$^{7}$}
\authorrunning{McDermid et al. ``Young KDCs in Early-Type Galaxies''}
\institute{Leiden Observatory, Postbus 9513, 2300 RA Leiden, The Netherlands
\texttt{mcdermid@strw.leidenuniv.nl}
\and CRAL-Observatoire, 9 Avenue Charles-Andr{\'e}, 69230 Saint-Genis-Laval, France
\and UC Berkeley Department of Astronomy, Berkeley, CA 94720, USA
\and Denys Wilkinson Building, University of Oxford, Keble Road, Oxford, UK
\and STECF/ESO, Garching, Germany
\and Kapteyn Institute, Postbus 800, 9700 AV Groningen, The Netherlands
\and Centre for Astrophysics Research, University of Hertfordshire, Hatfield, UK
}
%
%
\maketitle

\begin{abstract}
We present results from a series of follow-up observations of a sub-sample of the representative {\tt SAURON} survey elliptical (E) and lenticular (S0) galaxies using the {\tt OASIS} integral-field spectrograph. These observations focus on the central $10^{\prime\prime} \times 10^{\prime\prime}$, with roughly double the spatial resolution of the {\tt SAURON} observations. This increased spatial resolution reveals a number of interesting and previously unresolved features in the measured stellar kinematics and absorption-line strengths. We find that galaxies exhibiting the youngest {\it global} stellar populations (as measured with {\tt SAURON}) often contain a distinctly young {\it central} region (on scales of a few hundred parsec or less) compared to the rest of the galaxy. Moreover, these compact, young components are found to be mostly counter-rotating with respect to the rest of the galaxy. Given that there is no well-established reason for such young components to `prefer' counter- over co-rotation, this finding raises the following questions: How common are these small KDCs as a function of age? Why are there more young than old compact KDCs? Where are the equivalent co-rotating components? We explore these questions using simple simulated velocity fields and stellar population models, and find that the fading of the young component as it evolves, coupled with the fact that counter-rotating components are more easily detected in the velocity field, may help explain the observed trends.
\end{abstract}

\section{Young Kinematically Decoupled Components}

Since the first applications of absorption line indices as a diagnostic tool for studying stellar populations in early-type galaxies, it was found that some of these evolved and dynamically relaxed objects contain a non-negligible population of young stars (e.g. \cite{worthey94}). The distribution of this `frosting' of young stars within a galaxy was then largely uncertain, given the difficulties in measuring spatially-resolved absorption-line strengths. With the advent of integral-field spectroscopy, it is however now possible to obtain high-quality `maps' of absorption-line strength distributions within galaxies (e.g. see \cite{kuntschner06}, these proceedings), and in turn, maps of luminosity-weighted age, metallicity and abundance ratio by applying modern stellar population models.

%
%
\begin{figure}[t]
\centering
\includegraphics[width=10cm]{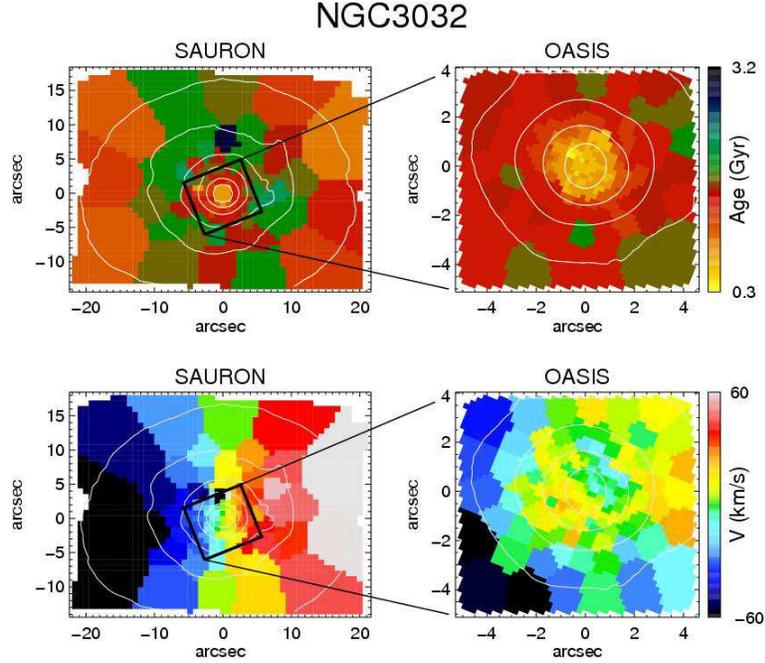}
\caption{{\it Top row:} Maps of mean stellar age, derived from line-strength measurements. {\it Bottom row:} Maps of mean stellar rotation velocity. The maps are derived from observations using {\tt SAURON} (left) and {\tt OASIS} (right), where the box indicates the {\tt OASIS} field.}
\label{fig:mcdermid_fig1}
\end{figure}

From such stellar population maps, it is evident that many galaxies showing globally young ages also tend to show centrally concentrated young components. Figure \ref{fig:mcdermid_fig1} shows a clear example of this in the dusty S0 galaxy NGC\,3032. The top row shows the mean stellar age, derived from maps of line-strength indices (H$\beta$, Fe5015, Mg{\it b}, and Fe5270) observed with {\tt SAURON} (left) and {\tt OASIS} (right), using the single-burst stellar population (SSP) models of \cite{thomas03}, where the SSP which best reproduced the multiple observed indices was found at each position. The galaxy shows rather young ages across the whole field, but shows a distinct decrease in age in the central $1^{\prime\prime}$ radius. The bottom panels of Figure \ref{fig:mcdermid_fig1} show the corresponding velocity maps for this galaxy from the same two instruments. The increased spatial resolution of the {\tt OASIS} observations reveals a small (3.5$^{\prime\prime} \equiv 370$~pc diameter) kinematically decoupled component (KDC), which coincides with the location of the central young population.

\begin{figure}[t]
\centering
\includegraphics[width=12cm]{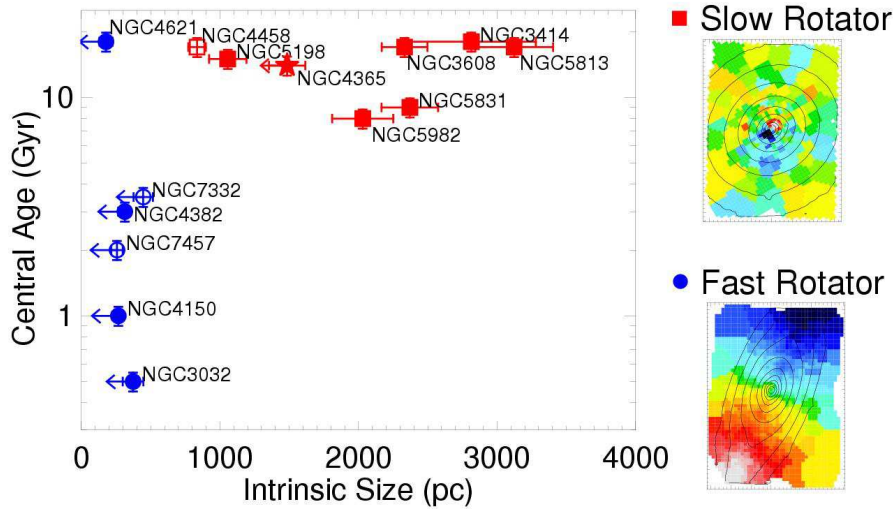}
\caption{Plot of central stellar age (where the KDC is assumed to dominate) against intrinsic KDC diameter (measured from velocity maps). Symbols indicate galaxies with both {\tt OASIS} and {\tt SAURON} data (filled) or only {\tt SAURON} (open); and whether the galaxy is a slow (square) or fast (circle) rotator (illustrated by the right-hand maps). Arrows indicate upper limits on the KDC diameter estimate. NGC\,4365 is included (star symbol) from \cite{davies01}.}
\label{fig:mcdermid_fig2}       
\end{figure}

Combining the available {\tt SAURON} and {\tt OASIS} data, we find a number of galaxies within our sample which contain similar small ($<$ few hundred parsec), young ($<$ 5 Gyr) KDCs, several of which are only resolved with {\tt OASIS}. Figure \ref{fig:mcdermid_fig2} shows the distribution of KDC size (estimated from the velocity maps) against the mean luminosity-weighted age measured within the central arcsecond of the galaxy (where light from the KDC is assumed to dominate) for all E/S0 galaxies in the {\tt SAURON} sample which show a clear KDC (i.e. neglecting co-rotating components). From this figure, we see that intrinsically large ($>$ 1kpc) KDCs tend to be rather old, and the compact KDCs tend to be young, although they cover a range in age. Moreover, the large KDCs are found exclusively in galaxies showing low global angular momentum, which we term `slow rotators'; the small KDCs on the other hand inhabit `fast rotators', which show significant net rotation. The young KDCs are also generally counter-rotating systems, sharing almost the same rotation axis as the outer parts (within $\sim 10^\circ$ in most cases).

%
%

\section{Interpretation}

\begin{figure}[t]
\centering
\includegraphics[width=7cm]{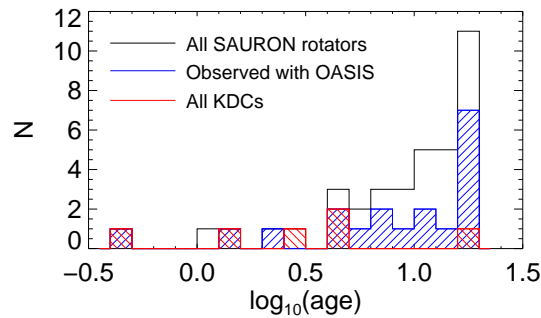}
\caption{Histogram of ages for the fast rotator galaxies from the {\tt SAURON} survey. From the complete sample (black line), we indicate those observed so far with {\tt OASIS} (blue), and those with a clear KDC (red).}
\label{fig:mcdermid_fig3}       
\end{figure}

Why are the small KDCs mostly young? Figure \ref{fig:mcdermid_fig3} shows that, of the fast rotating galaxies in the {\tt SAURON} sample, five of the seven youngest objects ($< 5$~Gyr) show a counter-rotating core, and one of these has yet to be observed with high spatial resolution, and so may yet reveal a KDC. Only one of the remaining thirteen objects older than 5~Gyr shows a detectable KDC.

Is the counter-rotation significant in producing the young stars? There are also galaxies showing young/intermediate-aged global and central populations, but which don't show strong substructure in their velocity fields. Are these `single component' systems fundamentally different?

%
%

\begin{figure}[t]
\centering
\includegraphics[width=12cm]{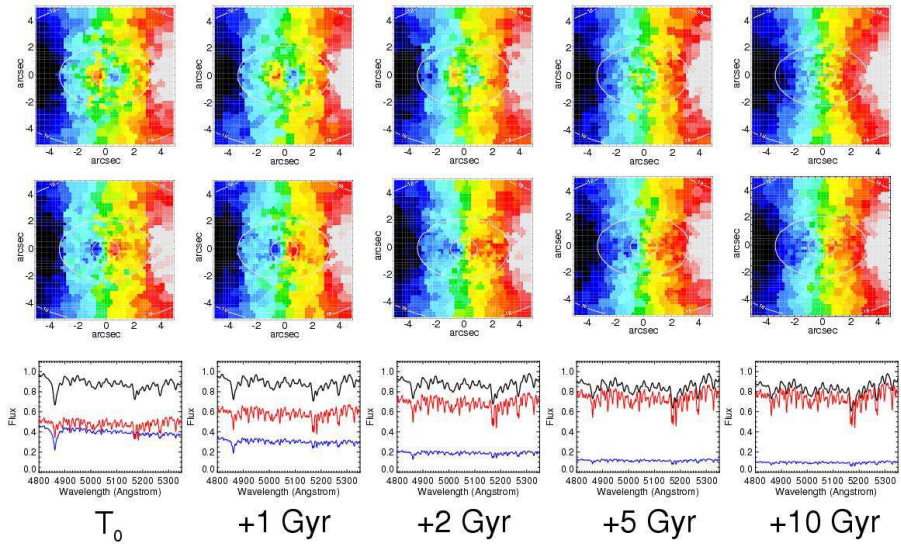}
\caption{Model velocity fields for an evolving counter-rotating (top row) and co-rotating (middle row) young subcomponent. The spectrum of the young decoupled component (blue) and older main component (red) are shown on the bottom row, indicating there luminosity-weighted contribution to the observed spectrum (black).}
\label{fig:mcdermid_fig4}       
\end{figure}

Figure \ref{fig:mcdermid_fig4} shows a possible answer to these questions. Using the measured central line-strengths from one of our young KDC galaxies, we constrain the amount of mass of young stars that can be added to a background `base' population (assumed to be that of the main body of the galaxy) within the central aperture. Taking the example of NGC\,4150, we add 8\% {\it by mass} of a 0.5~Gyr population on a 5~Gyr base population, to give a combined H$\beta$ absorption strength of $\sim 4$~\AA. We simulate a two-component velocity field using a Fourier expansion technique similar to the `kinemetry' method of \cite{krajnovic06}. We assign these velocities at each position to SSP model spectra of \cite{bruzual03}, assigning the young population to the KDC component, and make the mass-weighted combination of spectra. The simulated data cube is then spatially binned using a Voronoi tessellation \cite{cappellari03}, and the kinematics were extracted using the penalised pixel fitting technique of \cite{cappellari04}.

%
%

The result is a realistic-looking KDC velocity field. We then hold the mass fraction fixed, and `evolve' the populations in step. As the KDC population ages, its mass-to-light ratio increases, resulting in a dimming of the KDC stars. The effect is to `fade' the KDC into the background rotation field, as the luminosity-weighted contribution becomes less significant. After 5~Gyr, the KDC is barely visible. This helps explain the apparent lack of intermediate and old aged small KDCs.

The importance of counter-rotation is harder to address. The middle row of Figure \ref{fig:mcdermid_fig4} uses the same kinematic components and populations as the top row, but in this case the subcomponent is co-rotating. The impact of the co-rotating component on the total observed velocity field is clearly more subtle, and the component becomes difficult to separate from the background field after only $\sim 2$~Gyr. For this reason, the intrinsic distribution of co- and counter-rotating central young components could be rather similar, but since the co-rotating cases can only be clearly identified at young ages, our sample of young galaxies is currently too small to answer this question satisfactorily.

%
%
%


\printindex
\end{document}